# Direct Observation of Magnetic Gradient in Co/Pd Pressure-Graded Media

B. J. Kirby[1,a)], S. M. Watson[1], J. E. Davies[2], G. T. Zimanyi[3], Kai Liu[3], R. D. Shull[2], and J. A. Borchers[1]

[1]*Center for Neutron Research, National Institute of Standards and Technology, Gaithersburg, Maryland, 20899, USA*

[2]*Metallurgy Division, National Institute of Standards and Technology, Gaithersburg, Maryland, 20899, USA*

[3]*Physics Department, University of California, Davis, California, 95616, USA*

ABSTRACT

Magnetometry and neutron scattering have been used to study the magnetic properties of pressure graded Co/Pd multilayers. The grading of the multilayer structure was done by varying the deposition pressure during sputtering of the samples. Magnetic depth profiling by polarized neutron reflectometry directly shows that for pressure-graded samples, the magnetization changes significantly from one pressure region to the next, while control samples sputtered at uniform pressure exhibit essentially uniform magnetic depth profiles. Complementary magnetometry results suggest that the observed graded magnetic profiles are due in part to a decrease in saturation magnetization for regions deposited at progressively higher pressure. Increased deposition pressure is shown to increase coercivity, and for graded samples, the absence of discrete steps in the hysteresis loops implies exchange coupling among regions deposited at different pressures.

[a)] Electronic mail: brian.kirby@nist.gov

In an effort to increase the storage density of magnetic media such as hard drives, much research work has been devoted to development of exchange spring and exchange coupled composite media.[1-5] Such media use soft magnetic layers to decrease the necessary write field, and exchange couple those layers to hard magnetic layers that provide thermal stability. Recent calculations by Suess, *et al*. have proposed that a gradual transition from soft to hard anisotropy can result in additional gains in writeability while preserving thermal stability.[6] While the advantages of "graded" media have been shown theoretically, real structures with the predicted properties are difficult to realize experimentally. In order to advance this technology, it is therefore important to synthesize and characterize basic structures based on well understood materials. One good example structure is the Co/TM (TM = Pd, Pt) multilayer stack, which exhibits strong perpendicular anisotropy and tunable magnetic properties. In-situ variation of growth properties such as layer thickness and sputtering pressure allow for such magnetic tuning throughout the multilayer stack. Specifically, increased growth pressure introduces increased disorder, which significantly raises the coercivity.[7] While grading of Co/TM multilayers in this way appears feasible, proper characterization of the resultant magnetic properties poses a second challenge. For example, the effects of exchange coupling on the magnetization reversal behavior within such a structure are not well understood. Further, it is important to characterize graded materials not only with techniques that probe the collective behavior (e.g. conventional magnetometry), but also with those sensitive to the properties of the individual components.

In this work, we have fabricated Co/Pd multilayers grown by varying the sputtering pressure during deposition, and studied them using conventional magnetometry and polarized neutron reflectometry. We observe that pressure grading produced samples with a graded magnetization as desired, and we observe evidence of exchange coupling among magnetically different regions.

For this study we examined [Co(0.4 nm)/Pd(0.6 nm)]$_{59}$/Co(0.4 nm) multilayer films. The samples were grown at room temperature by dc magnetron sputtering in a vacuum chamber with a base pressure of $10^{-6}$ Pa. Multilayer thin films of [Co(0.4 nm)/Pd(0.6 nm)]$_{59}$/Co(0.4 nm) were deposited onto Si substrates and a 20 nm Pd seed layer, and capped with 5 nm Pd. During deposition, the Ar sputtering gas pressure was varied between 0.7 Pa and 2.7 Pa in order to vary the interface roughness and grain size as a function of depth into the film. Note that the above thicknesses are approximate, as some variation was observed with pressure. Here we focus on four samples:

(1) Sputtered entirely at 2.7 Pa

(2) Sputtered entirely at 0.7 Pa

(3) Pd seed and first 30 Co/Pd bilayers sputtered at 0.7 Pa, all else sputtered at 2.7 Pa

(4) Pd seed and first 30 Co/Pd bilayers sputtered at 0.7 Pa, next 15 bilayers sputtered at 1.6 Pa, all else sputtered at 2.7 Pa

We refer to the four different samples as "high pressure", "low pressure", "two pressure" and "three pressure" respectively.

Polarized neutron reflectometry (PNR) measurements were conducted using the NG-1 Reflectometer[8] at the NIST Center for Neutron Research. PNR is a technique sensitive to the compositional and magnetic depth profiles of thin film samples.[9,10] For our measurements, a 4.75 Å wavelength neutron beam was polarized by a polarizing supermirror and Mezei spin-flipper to be alternately spin-up (+) or spin-down (-) relative to an applied magnetic field $H$, and was incident on the sample. Although the Co/Pd system has a strong perpendicular anisotropy, specular PNR measurements are wholly insensitive to any component of the sample magnetization normal to the sample surface. [11,12] Therefore, $H$ was applied so as to bend the sample magnetization into the hard in-plane direction. The two non spin-flip (NSF) polarization cross sections ($R^{++}$ and $R^{--}$), and the two spin-flip (SF) cross sections ($R^{+-}$ and $R^{-+}$) were measured as a function of scattering vector $Q$. The data were corrected for background, neutron polarization efficiency (typically < 97%), and beam footprint. SF scattering - which originates purely from the component of the in-plane sample magnetization - was found to be small for all measurements and was not considered in the analysis. The sample's depth dependent nuclear scattering length density $\varrho(z)$ (a function of the characteristic scattering strengths of different nuclei), and the component of the magnetization parallel to $H$, $M_{plane}(z)$, was determined by model fitting the NSF data using the GA_REFL software package.[13]

Figure 1 shows the fitted NSF data, and the models used to fit the data for measurements taken at room temperature in $\mu_0 H = 0.7$ T, which is not sufficient to saturate $M_{plane}$. For convenience, the fitted data are plotted as spin asymmetry $A = (R^{++} - R^{--})/(R^{++} + R^{--})$. For simplicity, we modeled the 119 individual layers in the [Co/Pd] repeating bilayer as only three sections corresponding to the three different regions in the three pressure sample.[14] This three region scheme was applied to the fitting of all four samples, so as to have a fair comparison of the depth profiles. Evident from the nuclear profiles used to fit the data (Fig. 1 center column), we could detect no variation in $\varrho(z)$ throughout the Co/Pd stack for uniform or pressure-graded samples. However, for the high pressure sample the Co/Pd stack (60 nm) is noticeably thicker than it is for the low pressure sample (45 nm), implying some change in composition with increasing pressure.

The profiles of the in-plane component of the magnetization $M_{plane}$ (Fig. 1 far right) clearly show the effects of pressure grading. For the single pressure samples, $M_{plane}$ is fairly uniform, and is significantly smaller for the high pressure sample (180 kA/m)[15] than it is for the low pressure sample (330 kA/m). This pressure dependence is maintained for the individual sections of the two and three pressure samples, which feature very non-uniform magnetic depth profiles.[16] These results unambiguously confirm that varying the pressure during deposition did induce an actual *magnetic* gradient in the samples.

Higher field superconducting quantum interference device (SQUID) magnetometry measurements (5 T in-plane loops, not shown) provide the total moment per unit area for small pieces of the samples examined by PNR. Normalization of these values by the sample area and the total Co/Pd thickness obtained from PNR show that the low pressure sample has a much higher saturation magnetization (550 kA/m) than does the high pressure sample (400 kA/m), similar to what has been observed for Co/Pt films.[17] Thus, we conclude that the observed magnetization gradient observed from PNR (Fig. 1) is due in part, to a decrease in the total moment density for layers deposited at higher pressure. It is likely that the multi-pressure samples also exhibit a depth-dependent anisotropy that contributes to the observed magnetization gradient, but isolating this component requires precise measurement of the saturation magnetization of each individual layer, which is a topic for future investigations. Although we cannot clearly distinguish between the individual contributions of anisotropy and total moment, these measurements clearly show that the pressure graded samples exhibit a gradient in magnetic properties.

The results of additional PNR measurements taken after reducing $\mu_0 H$ from 0.7 T are shown in Figure 2. If exchange coupling between the magnetically different pressure regions is very strong, the field response of each region in a multi-pressure sample might be expected to differ relative to the field response of the corresponding single pressure sample. However, the field-dependent magnetization of regions deposited at the same pressure for different samples show little difference.

It is likely that this lack of coupling evidence from PNR is due to the small field range studied, as vibrating sample magnetometry measurements along the perpendicular easy axis do imply coupling. Figure 3 shows the results of these measurements for the four samples. The effect of increased pressure is evident as the coercivity is significantly higher for the high pressure sample than for the low pressure sample. The coercivities of the multi-pressure samples fall in between those of the single pressure samples, but exhibit no "steps" in the hysteresis loops indicative of sharp switching of individual pressure regions. Instead, gradual transitions - characteristic of spring magnets,[4] - are observed for the multi-pressure samples, strongly suggesting coupling among the magnetically different pressure regions.

It is interesting to note that the three pressure sample exhibits a larger coercive field than the two pressure sample, as additional steps in the graded structure should make for a softer magnet.[6] This is likely due to a competing effect caused by a change of the magnetization reversal mechanism. Previous studies on similar Co/Pt films have shown that with increasing sputtering pressure, the coercivity increases [17] as the reversal changes from domain nucleation and wall motion [7] to domain wall pinning and magnetization rotation. In our samples, substituting part of the layers grown at low pressure with those grown at higher pressure would also contribute to a higher coercivity.

In summary, we have prepared graded Co/Pd multilayer samples, and unambiguously confirmed that the pressure grading induces a corresponding depth-dependent magnetization. In addition, our results provide strong evidence of exchange coupling among the different magnetic regions in the multilayers. The magnetic behavior of our pressure-graded multilayers is thus consistent with many aspects of the ideal, proposed system, providing a possible approach for achieving graded magnetization in actual media applications.

Work at UCD has been supported by CITRIS.

[14] For reflectometry, the resolvable length scale roughly scales as $2\pi$ divided by the maximum $Q$ value, and for a superlattice, the $Q$ position of the first Bragg peak appears at $2\pi$ divided by the layer repeat thickness. Given the available $Q$ range for these measurements, our data are totally insensitive to

the contrast between the sub-nm Co and Pd layers, and do not reach far enough to access a superlattice Bragg peak.

FIGURE CAPTIONS

Figure 1: Fitted PNR data, and the resulting nuclear and magnetic depth profiles measured at 0.7 T. Increased deposition pressure results in a thicker Co/Pd stack with reduced magnetization. Single pressure samples (a and b) display fairly constant magnetic profiles, while those of the multi-pressure samples (c and d) are non-uniform. Error bars correspond to +/- 1 sigma.

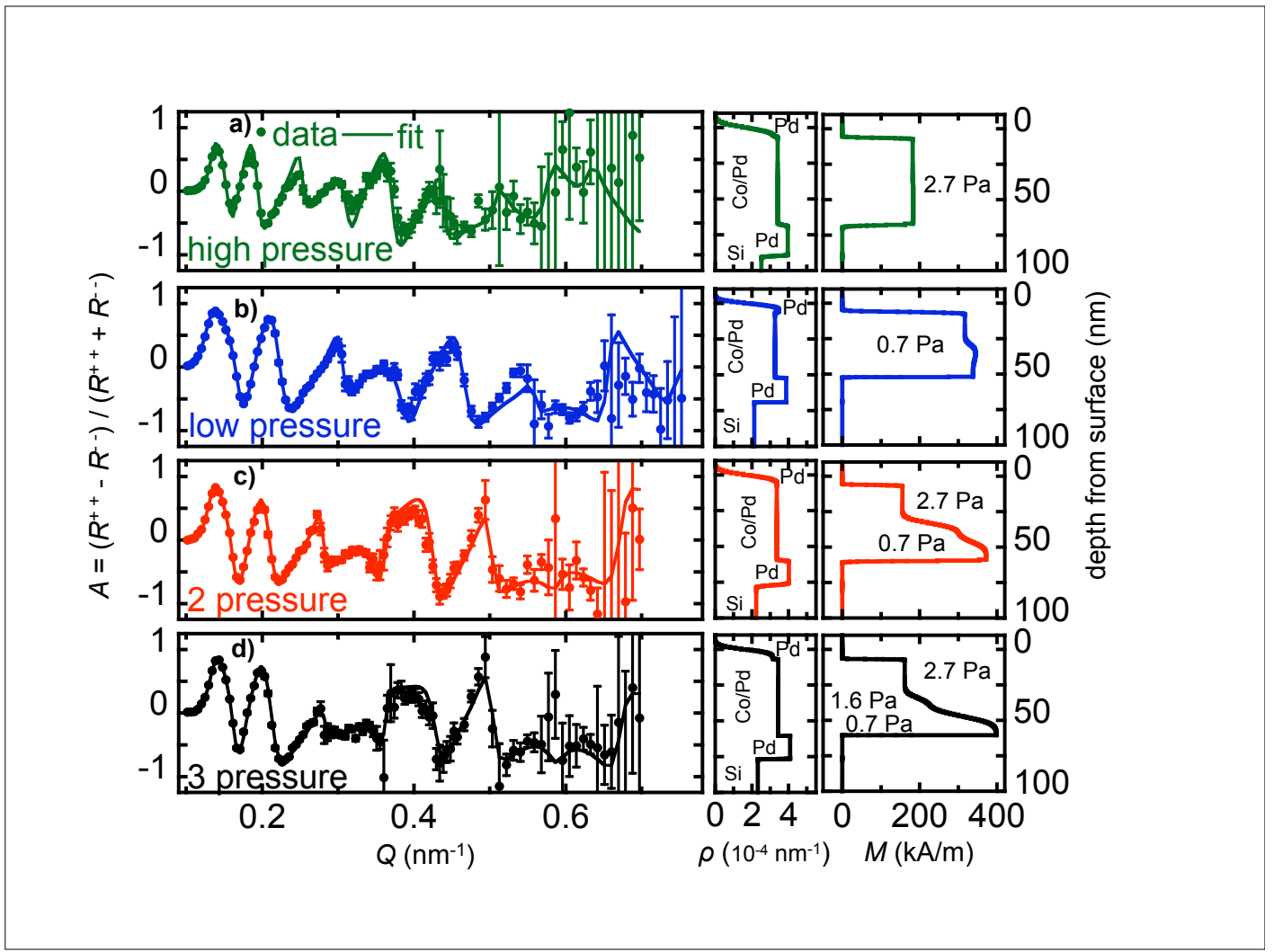

Figure 2: Field-dependent magnetizations of regions deposited at the same pressure for different samples, as measured with PNR.

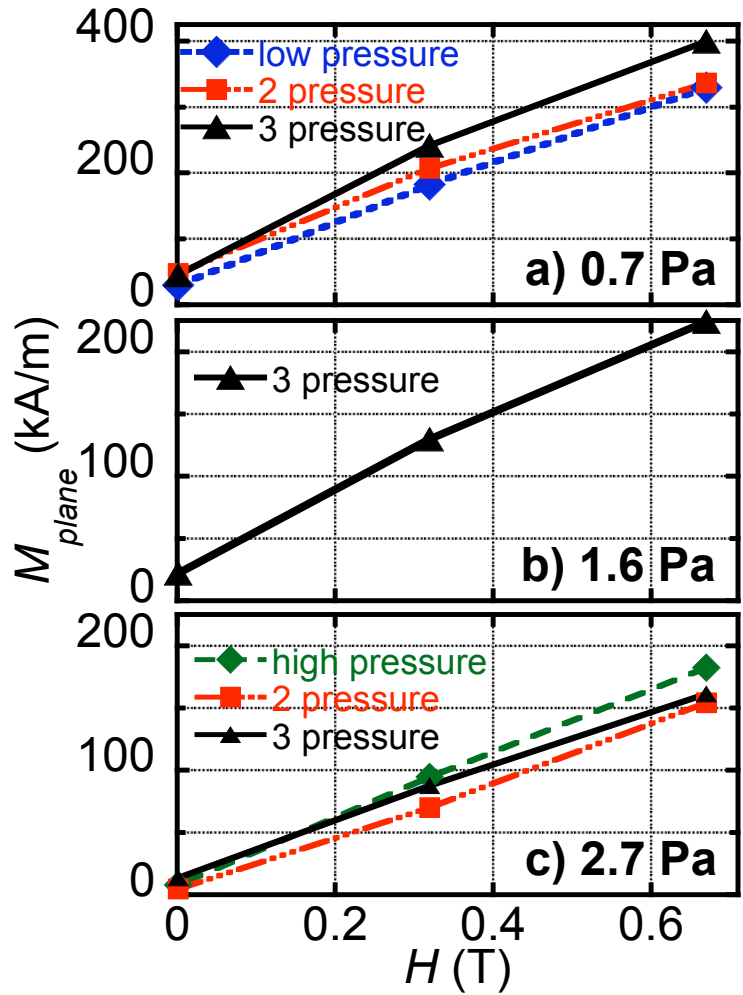

Figure 3: Field-dependent easy axis (perpendicular-to-plane) magnetizations, as measured with vibrating sample magnetometry.

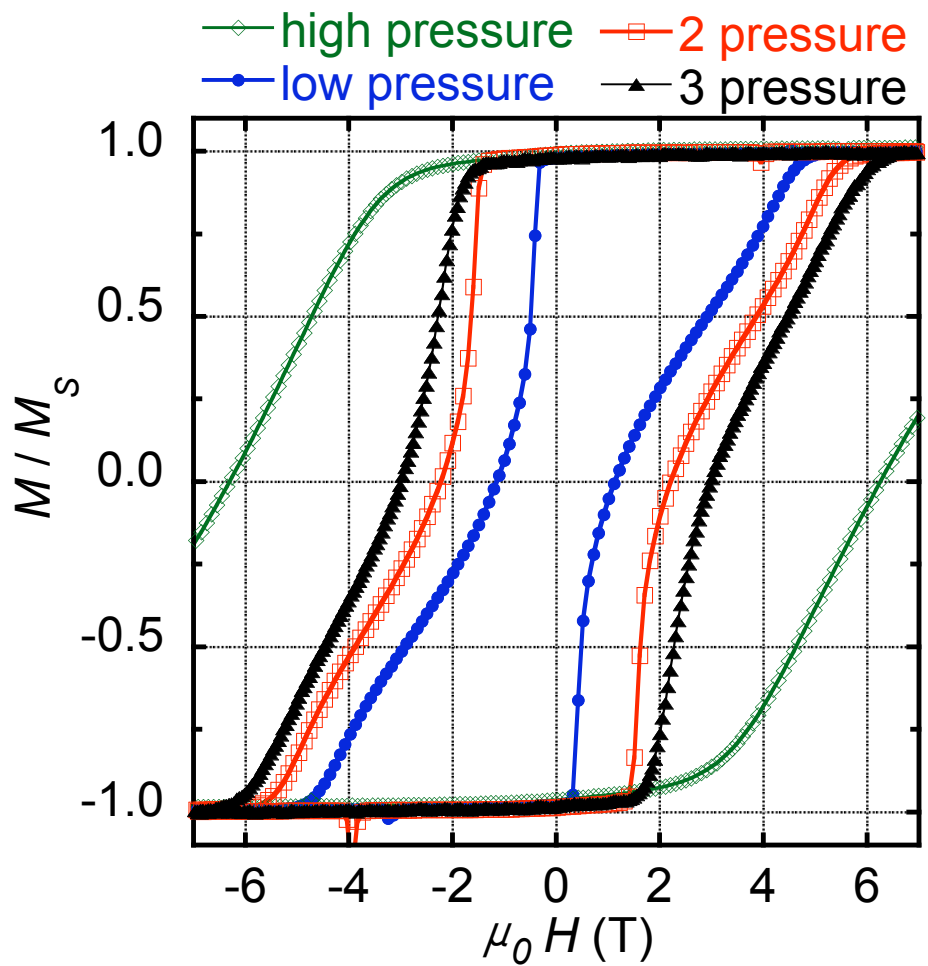